\documentclass[twocolumn,showpacs,aps,amsmath,amssymb]{revtex4}
\usepackage{graphicx}

\def\g5{\gamma_{5}}
\def\ga{\gamma}

\def\e{\epsilon}

\def\be{\begin{eqnarray}}
\def\ed{\end{eqnarray}}

\def\non{\nonumber}

\def\UP{\cal U}
\def\la{\langle}
\def\ra{\rangle}

\begin{document}
\title{\Large \bf
Unparticle physics on direct CP violation }
\date{\today}

\author{ \bf  Chuan-Hung Chen$^{1,2}$\footnote{Email:
physchen@mail.ncku.edu.tw} and Chao-Qiang
Geng$^{3,4}$\footnote{Email: geng@phys.nthu.edu.tw}
 }

\affiliation{ $^{1}$Department of Physics, National Cheng-Kung
University, Tainan 701, Taiwan \\
$^{2}$National Center for Theoretical Sciences, Hsinchu 300, Taiwan\\
$^{3}$Department of Physics, National Tsing-Hua University, Hsinchu
300, Taiwan  \\
$^{4}$Theory Group, TRIUMF, 4004 Wesbrook Mall, Vancouver, B.C V6T
2A3, Canada
 }

\begin{abstract}
The effects of the peculiar CP conserving phases in unparticle
propagators are explored. We find that the phases have a great
impact on  CP-violation. We adopt the decays $B_{d}\to \pi^{+}
\pi^{-}$ and $B_{d}\to \ell^{-}\ell^{+}$ as the illustrators to
demonstrate the influences of these phases on the direct CP
asymmetries. In particular, we emphasize that unparticle physics is
the only model suggested to date that could give the direct CP
asymmetries in $B_d\to \ell^{-} \ell^{+}$ as large as 15$\%$. We
also point out that the unparticle phases could be probed in $B\to
K^* \ell^{+} \ell^{-}$ decays by using T-odd correlations.

\end{abstract}
\maketitle

Recently, it has been argued in Refs. \cite{Georgi1,Georgi2} that a
scale invariant sector with  scale dimension $d_{\UP}$ is associated
with ``unparticle stuff'' which looks like a non-integral number
$d_{\UP}$ of invisible particles. The production of  unparticles
might be detectable by measuring the missing energy and momentum
distributions in various processes \cite{Georgi1,Georgi2}. The
unparticle physics phenomenology is further implemented in Refs.
\cite{TC,Luo}. Although it is still unclear how far the unparticle
physics can be carried out, there exist many possible interesting
experimental tests on unparticles. In particular, it has been
illustrated in Ref. \cite{Georgi2} that the peculiar CP conserving
phases associated with the unparticle propagators in the time-like
region lead to unusual CP conserving interference patterns between
the time-like unparticle exchange amplitudes and the standard model
(SM) amplitudes in $e^{+}e^{-}\to\mu^{+}\mu^{-}$.
 The effect of the virtual unparticle propagation  has
also been considered in Ref. \cite{TC}. It should be interesting to
ask if there are other odd effects due to this type of the phases.
In this study, we will examine
the effects of these
phases on CP violation.

In a decay process, the direct CP asymmetry (CPA) is defined by
 \be
{\cal A}_{CP}&\equiv&{\bar\Gamma -\Gamma\over \bar \Gamma+\Gamma}
 \ed
 where $\Gamma$ ($\bar{\Gamma}$) is the partial decay rate of the
(CP-conjugate) process. It is well known that
 \be
{\cal A}_{CP}&\propto&\sin\theta_{w}\sin\theta_{st}\,,
 \ed
where $\theta_{w}$ and $\theta_{st}$,
 are CP violating (CPV) and CP conserving (CPC) phases,
 so called weak and strong phases, respectively. In the SM, the
weak
CPV phase is the unique phase in the $3\times
3$ Cabibbo-Kobayashi-Maskawa (CKM) matrix \cite{CKM}, which appears
in the CKM matrix elements of $V_{td}$ and $V_{ub}$
\cite{Wolfenstein}, expressed by $V_{td}=|V_{td}|e^{-i \beta}$ and
$V_{ub}=|V_{ub}|e^{-i\ga}$ \cite{PDG06}. With $10^{8}$ $B\bar B$
pairs produced at  B factories, the measurement on $\sin2\beta$
through the Golden mode of $B_d\to J/\Psi K_s$ is determined to be
$0.73 \pm 0.04$ \cite{PDG06}. It has no doubt that the mixing
induced CPA in the B system is dominated by the CKM phase without
the need of a strong phase.

To probe the phase effects, one can also use some T-odd correlations
such as the well known
triple spin-momentum product
correlation
in a three-body decay \cite{TPC}.
As the T-odd effects of the correlations are proportional to
$\sin\theta_{w}\cos\theta_{st}$ as well as
$\cos\theta_{w}\sin\theta_{st}$,
they can be generated by either CPV or CPC phase. As a result,
if the CPV phase in the process
vanishes, i.e. $\theta_{w}=0$, one can utilize the T-odd correlations
to test the CPC phase of $\theta_{st}$.

In this Letter, we will use $B$ decay processes to illustrate
the impact of the CPC phases in the unparticle propagators on CP violation.
Explicitly,
we will show that without the strong
QCD phases, the
large direct CPA in $B_d\to \pi^- \pi^+$ could be generated when
the unparticle physics is introduced.
We will also demonstrate that the unparticle phase
could create non-zero direct CPAs in $B_d \to \ell^{-} \ell^{+}$
decays, which are vanishing small in most of other CP violating
models due to the lack of strong phases. Furthermore, we consider
T-odd observables in the decays
of
$B\rightarrow K^{*} \ell^{+} \ell^{-}$ ($\ell=e,\, \mu$)
to reveal the unparticle phases since there are
 T-odd effects due to the zero CPV weak phase for the $b\to s$ transition in
the SM.

We start with the propagator of a scalar (vector) unparticle, given
by \cite{Georgi1,Georgi2}
  \be
  \int d^4x e^{i p\cdot x} \la 0| T\left( O^{(\mu)}_{\UP}(x)
O^{(\nu)}_{\UP}(0)\right) |0 \ra
 = i\Delta_{\UP}^{S(V)}(p^2)\,e^{-i\phi_{\UP}}
\label{eq:uprop}
 \ed
 with
 \be
  \Delta_{\UP}^{S}(p^2)&=& \frac{A_{d_{\UP}}}{2\sin(d_{\UP}
\pi)} \frac{1}{\left(p^2
+i\e\right)^{2-d_{\UP}}}\,,\non\\
  \Delta_{\UP}^{V}(p^2)&=& \frac{A_{d_{\UP}}}{2\sin(d_{\UP}
\pi)} \frac{ -g^{\mu \nu} + p^{\mu} p^{\nu}/p^2 }{\left(p^2
+i\e\right)^{2-d_{\UP}}}\,,
  \ed
   where $\phi_{\UP}=(d_{\UP}-2)\pi$,
$O^{(\mu)}_{\UP}$ is the scalar (vector) unparticle operator and \be
A_{d_{\UP}}&=&  \frac{16 \pi^{5/2}}{(2\pi)^{2d_{\UP}}}
\frac{\Gamma(d_{\UP}+1/2)}{\Gamma(d_{\UP}-1) \Gamma(2 d_{\UP})}\,.
\label{eq:pro_up}
 \ed
Note that in Eq. (\ref{eq:uprop}) the phase factor arises from
$(-1)^{d_{\UP}-2}=e^{-i\pi(d_{\UP}-2)}$ and the vector operator is
assumed to satisfy the transverse condition $\partial_{\mu}
O^{\mu}_{\UP}=0$.

%
To study the unparticle physics, we write the effective interactions
for unparticle operators coupling to leptons and quarks to be
\cite{Georgi1,Georgi2}
 \be
&&  \frac{c^{\ell}_{A}}{\Lambda^{d_{\UP}-1}_{\UP}} \bar \ell
\ga_{\mu} \ga_5 \ell O^{\mu}_{\UP} +
\frac{c^{\ell}_{P}}{\Lambda^{d_{\UP}}_{\UP}} \bar \ell \ga_{\mu}
\ga_5 \ell \partial^{\mu} O_{\UP}+ \non\\
&& \frac{c^{q' q}_{VA}}{\Lambda^{d_{\UP}-1}_{\UP}} (\bar q' q)_{V-A}
O^{\mu}_{\UP} + \frac{c^{q' q}_{SP}}{\Lambda^{d_{\UP}}_{\UP}} (\bar
q' q)_{V-A} \partial_{\mu} O_{\UP}
 \label{eq:heff}
 \ed
where $(\bar q' q)_{V-A}=\bar q' \ga_{\mu} (1-\ga_5) q$, $c$'s are
the dimensionless parameters and $\Lambda_{\UP}$ is the energy scale
below which the scale invariant unparticle fields emerge. For
simplicity, in Eq. (\ref{eq:heff}) we have only chosen the relevant
interacting structures for $B_d\to \pi^{-} \pi^{+}$, $B_d\to
\ell^{-} \ell^{+}$ and
$B\rightarrow K^{*} \ell^{+} \ell^{-}$ and we have set that all $c$'s are real numbers,
as the general case is not the issue here.

In terms of the
factorization approach, the decay amplitude for $B_d\to \pi^{-}
\pi^{+}$ is expressed by
 \be
A(B_d\to \pi^- \pi^+)&=& \frac{G_F}{\sqrt{2}} V^{*}_{ub} V_{ud}a_1
m^2_B f_{\pi} F^{B\pi}_{0}(m^2_{\pi})\non\\
&& \times
 \left[1+ \chi_{\pi \pi} e^{-i \phi_{\UP}} e^{-i\ga}
\right]
 \ed
 with
 \be
\chi_{\pi\pi}&=& \frac{8}{g^2 a_1 N_c}\frac{c^{bd}_{VA}
c^{uu}_{VA}}{ |V^{*}_{ub} V_{ud}|} \frac{A_{d_{\UP}}}{2\sin d_{\UP}
\pi} \frac{m^2_W}{p^2}\left(\frac{p^2}{\Lambda^2_{\UP}}
\right)^{d_{\UP}-1}\,, \non
 \ed
where $a_1=C_2+C_1/N_c$ is the effective Wilson coefficient
\cite{BBL} with $N_{c}$ being the color number,
$p^{2}\sim m_{B}\bar{\Lambda}$ with $\bar{\Lambda}=m_{B}-m_{b}$
and $f_{\pi}$ and
$F^{B\pi}_{0}$ denote the pion decay constant and $B\to \pi$ form
factor, respectively. 
  We note that
the scalar unparticle contributions have been ignored as they are
suppressed by $m^2_{B}/\Lambda^2_{\UP}$.
 Subsequently, the
BR and direct CPA are found to be
 \be
 {\cal B}(\pi^{-} \pi^{+} ) &=& {\cal B}^{SM}_0\left( 1+
 \chi^2_{\pi\pi}+2 \chi_{\pi\pi} \cos d_{\UP}\pi \cos\ga
 \right),\non\\
 {\cal A}_{CP}&=& \frac{2 \chi_{\pi\pi} \sin d_{\UP}\pi \sin\ga}{1+
 \chi^2_{\pi\pi}+2 \chi_{\pi\pi} \cos d_{\UP}\pi \cos\ga}\,,
 \ed
where ${\cal B}^{SM}_0=\tau_{B_d} m^3_B G^2_F a^2_1 f^2_{\pi}
F^{B\pi}_{0}(m^2_{\pi}) |V^{*}_{ub} V_{ud}|^2/32\pi$ is the SM
contribution.

\begin{figure}[htbp]
\includegraphics*[width=2.5 in]{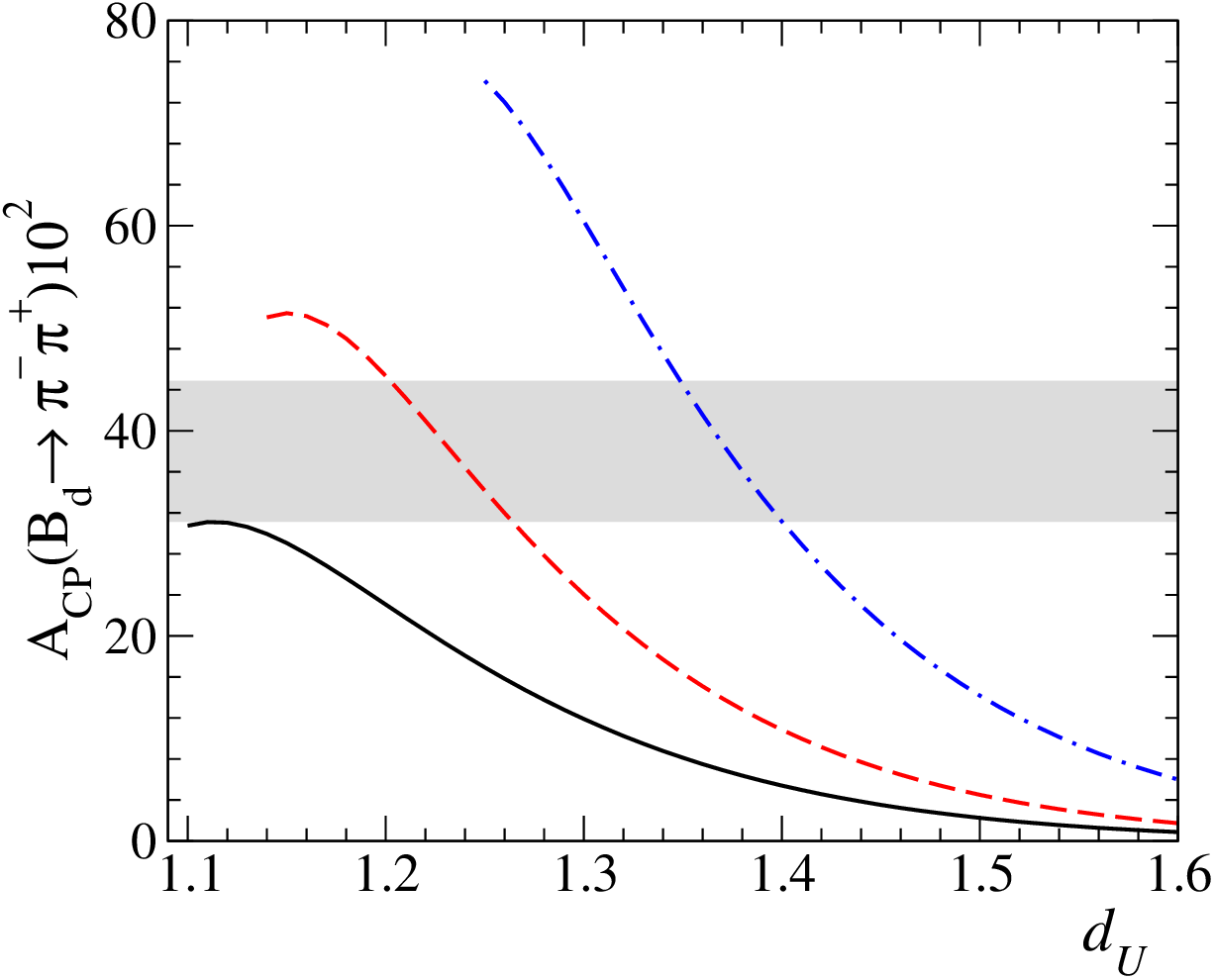}
\caption{Direct CP asymmetry
in $B_d\to \pi^- \pi^+$ as a function of $d_{\UP}$ with
$\Lambda_{\UP}=1$ TeV, where the sold, dashed and dash-dotted lines
correspond to $c^{uu}_{VA}=0.05$, $0.1$ and $0.5$, respectively, and
the band denotes the world average with 1$\sigma$ errors.}
 \label{fig:cpbpipi}
\end{figure}

The unknown parameter $c^{bd}_{VA}$ can be directly constrained by
the $B_d-\bar B_d$ mixing and the explicit relation is given by
 \be
\Delta M^{\UP_{\rm V}}_{B_d}& \approx &  \frac{f^2_B}{m_B}
\frac{A_{d_{\UP}}}{2\sin d_{\UP} \pi}  \left(
\frac{m^2_B}{\Lambda^2_{\UP}}\right)^{d_{\UP}-1}\left| c^{bd}_{VA}
\right|^2\,. \label{eq:dmB}
 \ed
 With $\Delta M_{B_d}^{\UP_{\rm V}}\leq 3 \times 10^{-13}$ GeV
\cite{PDG06}, $f_B=0.2$ GeV, $d_{\UP}=3/2$ and $\Lambda_{\UP}=1$
TeV, the order of magnitude for $c^{bd}_{VA}$ could be known as
$|c^{bd}_{VA}|\sim  10^{-4}$.
%
The direct CPA in $B_d\to \pi^- \pi^+$ as a function of the scale
dimension $d_{\UP}$ with $\Lambda_{\UP}=1$ TeV
 is presented in
Fig.~\ref{fig:cpbpipi}, where the solid, dashed, and dash-dotted
lines corresponds to $c^{uu}_{VA}=0.05$, $0.1$ and $0.5$,
respectively. The band in the figure denotes the world average
$0.38\pm 0.07$ \cite{HFAG}.
 From our results, we see clearly that the CPC phase in the unparticle physics
provides a mechanism to produce a large CPA for $B_d\to \pi^- \pi^+$
decay. However, we remark that both experimental
measurements \cite{BppExpt}
and theoretical
results in and beyond the SM \cite{JRS}
are still inconclusive.

Since the vector unparticle has only transverse degrees of freedom,
it has no effects on $B_d\to \ell^{-} \ell^{+}$ ($\ell=e,\mu$ and
$\tau$). However, the scalar unparticle could give contributions to
the decays. As a result, the decay amplitudes are found to be
 \be
A(B_{d}\to \ell^{+} \ell^-) &=& -i 2m_{\ell} m^2_B
 f_{B}
  \frac{c^{bd}_{SP}}{\Lambda_{\UP}^{d_{\UP}}}
  \frac{c^{\ell}_{P}}{\Lambda_{\UP}^{d_{\UP}}}
   \non\\
 &\times &
 \Delta^{S}_{\UP}(m^2_{B}) e^{-i\phi_{\UP}}\; \bar \ell \ga_5
  \ell\,;
 \ed
%
and the decay BRs for $B_d\to \ell^{-}
\ell^{+}$ are given by
 \be
\frac{1}{m^2_{\ell}}{\cal B}(B_d \to \ell^{+} \ell^{-})&=& \kappa_B
\left|Z^{B}_{SM} e^{-i\beta}+ Z^{B}_{\UP} e^{-i\phi_{\UP}}\right|^2
\non
 \ed
where
 \be
\kappa_B&=& \frac{1}{m^2_{\tau}} \frac{ \alpha^2 {\cal B}(B^+\to
\tau^{+} \nu_{\tau} )}{\pi^2\sin^4\theta_W}
\frac{\tau_{B_{d}}}{\tau_{B^+}}\,, \non \\
%
%
 Z^{B}_{\UP} &=& \frac{16\pi \sin^2\theta_W}{g^2 \alpha}
 \frac{ c^{b
d}_{SP}c^{\ell}_{P}}{|V_{ub}|}
\frac{A_{d_{\UP}}}{2\sin d_{\UP} \pi} \frac{m^2_W}{m^2_B} \left(
\frac{m^2_B}{\Lambda^2_{\UP}} \right)^{d_{\UP}}\,, \non
 \ed
and $Z^{B}_{SM}=|V^{*}_{tb} V_{td}| Y(x_t)/|V_{ub}|$ with
$x_t=m^2_t/m^2_{W}$. Here, we have used the measured $B^+\to \tau^+
\nu_{\tau}$ decay to remove the uncertainty from $f_B$. Due to
$m_{t}\gg m_{W}$, the function of $Y(x_{t})$ is simplified as
$Y(x_t)=0.315\, x_t^{0.78}$ \cite{BBL}. As the $B$ meson is much
heavier than leptons, we treat the leptons as massless particles.
The explicit $m_{\ell}$ comes from the equation of motion. In order
to make the discussions independent of lepton species, we will
concentrate on the quantity of ${\cal B}(B_d\to \ell^{-}
\ell^{+})/m^2_{\ell}$.

It is known that long-distance contributions which could introduce strong phases
are negligible in the pure leptonic B decays.
As a result,
 one cannot get  sizable direct CPAs in $B_d\to
\ell^{-} \ell^{+}$ decays in the SM
 although there exists
a weak phase $\beta$. Moreover, no matter how many
new weak CPV phases in new physics are injected, the conclusion cannot be
changed until  an intermediate state is involved, which carries an
exotic CPC phase. Amazingly, the peculiar phase associated with the
unparticle propagator is the unique phase that can generate non-zero
CPAs in $B_d\to \ell^{-} \ell^{+}$.

\begin{figure}[htbp]
\includegraphics*[width=3.1 in]{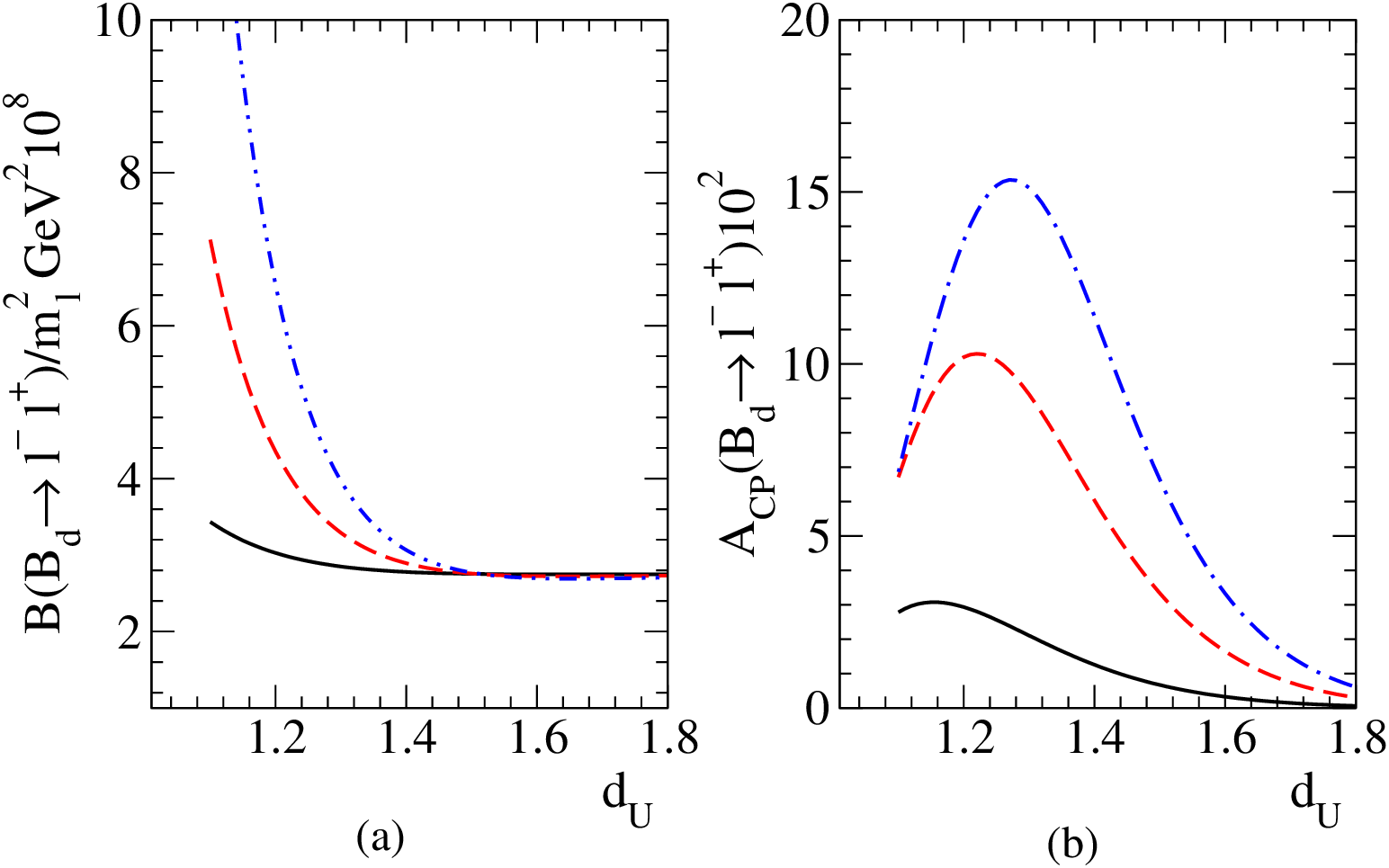}
\caption{(a) BR/$m^2_{\ell}$
and (b) direct CP
asymmetry
in $B_d\to \ell^- \ell^+$ as functions of $d_{\UP}$ with
$\Lambda_{\UP}=1$ TeV, where the sold, dashed and dash-dotted lines
correspond to $c^{\ell}_{P}=0.1$, $0.5$ and $1.0$, respectively. }
 \label{fig:bdll}
\end{figure}

Similar to $B_d\to \pi^{-} \pi^{+}$, $c^{bd}_{SP}$ could be
constrained by the $B_d-\bar B_d$ mixing, given by
\be \Delta M^{\UP_{S}}_{B_d}& \approx & \frac{5}{3}
\frac{f^2_B}{m_B} \frac{A_{d_{\UP}}}{2\sin d_{\UP} \pi} \left(
\frac{m^2_B}{\Lambda^2_{\UP}}\right)^{d_{\UP}}\left| c^{bd}_{SP}
\right|^2\,, \label{eq:dmBS}
 \ed
where the matrix element  $\la B_d | \bar b (1-\ga_5) d\, \bar b
(1-\ga_5) d| \bar B_d\ra \approx -5/6 f^2_B m_B$ has been used. With
${\cal B}(B^{+} \to \tau^{+} \nu_{\tau})=1.79\times 10^{-4}$
\cite{Belle_PRL97} and $\Lambda_{\UP}=1$ TeV, we present our
numerical results of the BR and the direct CPA in
Fig.~\ref{fig:bdll}, where the solid, dashed and dash-dotted lines
denote $c^{\ell}_P= 0.1$, $0.5$ and $1.0$, respectively.
 From Fig.~\ref{fig:bdll}a, we see that  when the scale dimension
 $d_{\UP}$ is less than
$1.4$, ${\cal B}(B_d\to \ell^{-} \ell^{+})/m^2_{\ell}$ becomes very
sensitive to  it.
The
merging flat lines for $d_{U}> 1.4$ correspond to the SM
contributions.  As shown in Fig.~\ref{fig:bdll}b,
it is very interesting to see that
${\cal A}_{CP}(B_d\to \ell^{-} \ell^{+})$ can be as large as
$15\%$. This is a very unique result in the unparticle physics, which
is different from most of other new physics.
If a large
deviation in BR for $B_d\to \ell^{-} \ell^{+}$ is found
experimentally, of course, it could be a signal for the existence of
new physics; but, it could not tell us what kind of new
physics involved. However, if a nonzero CPA in $B_d\to
\ell^{-} \ell^{+}$ is observed, the unparticle physics
definitely  plays the most important role.

Finally, we study the interesting T-odd operators
in $B\to K^* \ell^{+} \ell^{+}$ with polarized $K^*$
\cite{CG_PRD66}. With
the axial-vector interactions in Eq.~(\ref{eq:heff}), the
corresponding effective Hamiltonian for $b\to s \ell^{+} \ell^{-}$
with the SM effects is found to be
 \be
 {\cal H}_{\rm eff}&=& \frac{G_F\alpha_{em} V_{ts}V^{*}_{tb}}{\sqrt{2}
 \pi}\left[\left(C^{\rm eff}_{9}(\mu )\bar{s}\gamma _{\mu }P_{L}b \right.\right.\non\\
 &&\left.-\frac{2m_{b}}{%
 q^{2}}C_{7}(\mu )\bar{s}i\sigma _{\mu \nu }q^{\nu }P_{R}b \right)\bar{\ell}\gamma ^{\mu }\ell
 \non\\
&&\left. +C^{\UP}_{10}\bar{s}\gamma _{\mu }P_{L}b\; \bar{\ell}\gamma
^{\mu }\gamma
 _{5}\ell \right]\,,
  \ed
where $C^{\rm eff}_{9}$ and $C_{7}$ are the Wilson coefficients of
the SM \cite{BBL}. We note that $C^{\rm eff}_{9}$ includes the
$c\bar c$ resonant effects. The unparticle contribution only appears
in $C^{\UP}_{10}$ and is given by
  \be
 C^{\UP}_{10}=C_{10}+ \frac{16\pi c^{bs}_{VA} c^{\ell}_{A}}{g^2 \alpha_{em} V_{ts}V^*_{tb}}\frac{m^2_W}{q^2}
\left( \frac{q^2}{\Lambda^2_{\UP}}\right)^{d_{\UP}-1}\,.
 \ed
To study the T-odd spin-momentum correlation effects, we need to
consider the $K^*$ polarizations, {\it i.e.}, we have to consider
the decay chain $B\to K^* \ell^{+} \ell^{-}\to K\pi \ell^{+}
\ell^{-}$. Hence, in terms of the obtained Hamiltonian, the
differential decay rate for $B\to K\pi \ell^{+} \ell^{-} $
is obtain by
\begin{eqnarray}
&&\frac{d\Gamma }{d\cos \theta _{K}d\cos \theta _{\ell}d\phi dq^{2}}=\frac{%
3\alpha _{em}^{2}G_{F}^{2}\left| V_{ts}V^{*}_{tb}\right| ^{2}\left| \vec{p}%
\right| }{ 2^{14}\pi ^{6}m_{B}^{2}}  \non \\
&&\times Br(K^{*}\rightarrow K\pi )
\left\{\cdots + 2\sin2\theta _{K}\sin \theta _{\ell}
\sin \phi \right.\non\\
&&\left.\times \left(Im{\cal M}_{1}^{0} ({\cal M}_{2}^{+*}+{\cal M}_{2}^{-*})-Im({\cal M}_{1}^{+}+{\cal M}_{1}^{-})%
{\cal M}_{2}^{0*}\right) \right\}\,, \non\\\label{eq:diffrate}
\end{eqnarray}
where for simplicity we just display the terms associated with
dominant T-odd effect and the irrelevant terms are denoted by
$\cdots$, $\theta_{\ell (K)}$ is the polar angles of $\ell^{-} (K)$
in the rest frame of $\ell^{+} \ell^{-} (K^*)$, $\phi$ represents the
relative angle of the decaying plane between $K\pi$ and $\ell^{+}
\ell^{-}$, $q^2$ is the invariant mass of $\ell^{+} \ell^{-}$ and
$|\vec{p}
|=[((m_{B}^{2}+m_{K^{*}}^{2}-q^{2})/(2m_{b}))^{2}-m_{K^{*}}^{2}]^{1/2}$.
In Eq. (\ref{eq:diffrate}),
$ {\cal M}_{i}^{0}$ and ${\cal M}_{i}^{\pm }$ denote the
longitudinal and transverse polarizations of $K^{*}$
and their explicit expressions are given by
\begin{eqnarray*}
{\cal M}_{a\mu }^{(\lambda )} =if_{1}\varepsilon _{\mu \nu \alpha
\beta }\epsilon ^{*\nu }(\lambda )P^{\alpha }q^{\beta
}+f_{2}\epsilon _{\mu }^{*}(\lambda )+f_{3}\epsilon^{*}(\lambda)\cdot
qP_{\mu },
\end{eqnarray*}
respectively, where $P=p_{B}+p_{K^*}$, $q=p_{B}-p_{K^*}$ and
$a=1[2]$
while $f_{i}=h_{i}$ $[g_{i}]$ ($i=1$, $2$, $3$) with
\begin{eqnarray*}
 h_{1}& =&C_{9}^{\rm eff}(\mu )V(q^{2})-\frac{2m_{b}}{q^{2}} C_{7}(\mu )T(q^{2}),
 \\
h_{2(3)} &=& -C^{\rm eff}_{9}(\mu )A_{0(1)}(q^{2})+
\frac{2m_{b}}{q^{2}}
C_{7}(\mu )T_{0}(q^{2}), \\
g_{1}& =&C^{\UP}_{10}V(q^{2}),\ \ g_{2(3)}=-C^{\UP}_{10}A_{0(1)}(q^{2}),
\end{eqnarray*}
 and $V(q^2)$,
$A_{0(1)}(q^2)$ and $T_{0(1)}$ being the $B\to K^*$ form factors
\cite{CG_PRD66}.

\begin{figure}[hpbt]
\includegraphics*[width=2.5 in]{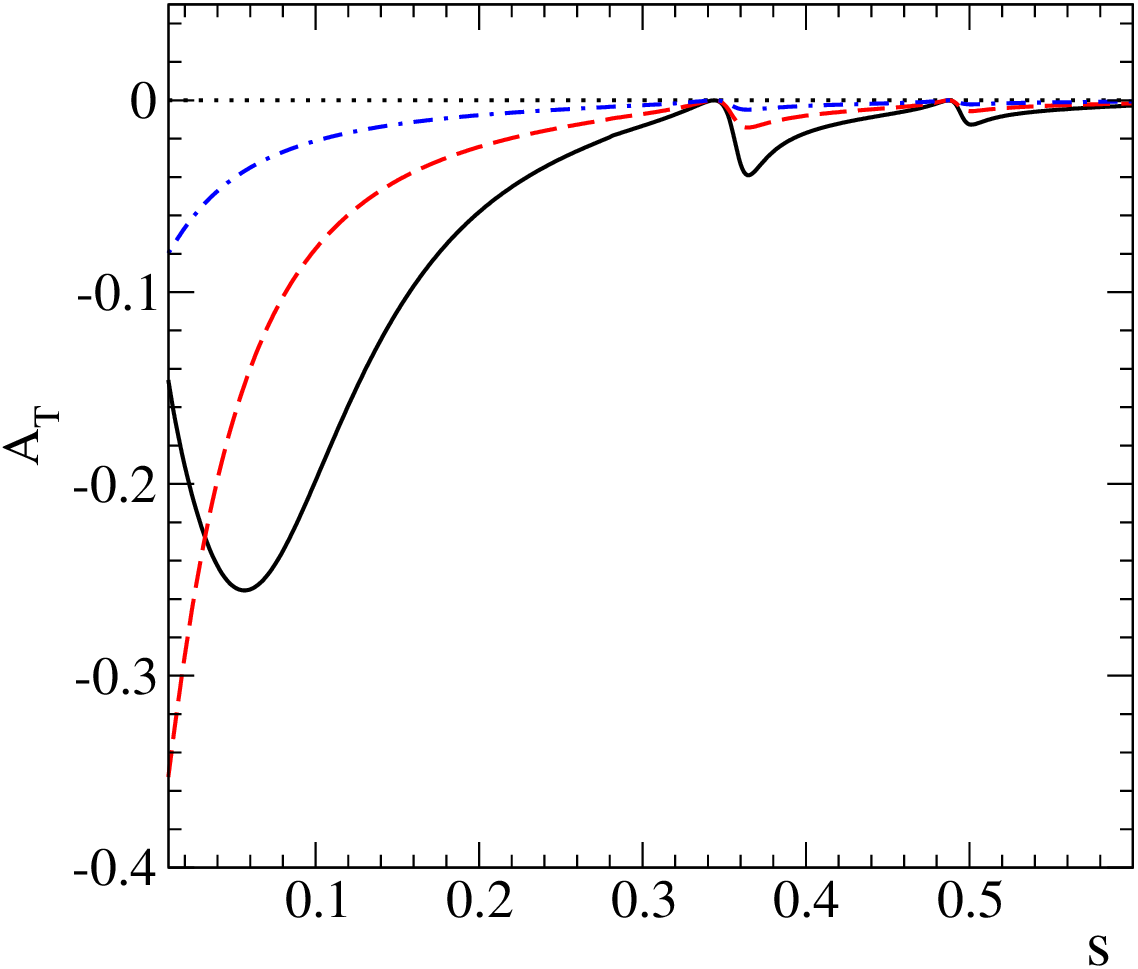}
\caption{ Statistical
significance $A_{T}$ of the T-odd observable in $B_{d}\to K^{*0} \ell^{+} \ell^{-}$ as
a function of $s=q^2/m^2_B$, where the solid, dashed and dot-dashed lines
correspond to $d_{\UP}=1.2$, $1.4$ and $1.6$ while the dotted line
is the SM result, respectively. }
 \label{fig:tv}
\end{figure}

 To illustrate the T-odd effects due to unparticles,
 we concentrate on the T-odd observable
defined by $ \left\langle {\cal O}_T\right\rangle =\int {\cal O}_T
d\Gamma$ \cite{CG_PRD66} where ${\cal O}_T$ is a T-odd five-momentum
correlation, given by
\begin{equation*}
{\cal O}_T=\frac{\left( \vec{p}_{B}\cdot \vec{p}_{K}\right) \left( \vec{p}%
_{B}\cdot (\vec{p}_{K}\times \vec{p}_{\ell^{+}})\right) }{\left| \vec{p}%
_{B}\right| ^{2}\left| \vec{p}_{K}\right| ^{2}\omega_{\ell^{+}} }%
\,
\end{equation*}
with $\omega_{\ell^{+}}=q\cdot p_{\ell^{+}}/\sqrt{q^2}$. In the
$K^{*}$ rest frame, we note that ${\cal O}_T=\cos \theta _{K}\sin
\theta _{K}\sin \theta _{\ell}\sin \phi $. The statistical
significance of the observable can be determined by
\begin{equation}
A_{T} (q^{2})={\frac{\int {\cal O}_Td\Gamma }{\sqrt{(\int d\Gamma
)(\int {\cal O}^{2}_T d\Gamma )}}}\,.  \label{ss}
\end{equation}
With the constraint of $\Delta m_{B_s}$ \cite{PDG06} on
$c^{bs}_{VA}$ and $c^{\ell}_{A}=5\times 10^{-4}$, the results
with unparticles on the T-odd observable
are displayed in Fig.~\ref{fig:tv}.
We note that the SM contribution vanishes as the associated CKM
matrix elements of $V_{ts}$ and $V_{tb}$ contain no CPV phase.
Moreover, it is interesting to point out that
even if we include the resonance effects, the T-odd effects also get
canceled in the SM. As seen from Fig.~\ref{fig:tv}, the unparticle
effects could make the statistical significance as large as $20\%$.
It is clear
if there is no any direct CP violating effect
in the $b\to s$ transition, {\it e.g.}, no $\sin2\beta$ difference
between
$B_d\to J/\Psi K_S$ and $B_d\to \phi K_S$, any signal
found in the T-odd obervables of $B\to K^* \ell^{+} \ell^{-}$ will
directly reveal unparticle phase effects.

In summary, we have illustrated that the peculiar CPC phases in the
unparticle propagators can play very important roles on the direct
CPAs in $B_{d}\to \pi^{-}\pi^{+}$ and  $B_{d}\to\ell^{-} \ell^{+}$  and
the T-odd observables in $B\to K^* \ell^{+} \ell^{-}$.
We have demonstrated
that
unparticle physics
is the only model suggested to date that
could give ${\cal A}_{CP}(B_d\to \ell^{-}
\ell^{+})$
as large as 15$\%$.
In addition, we have considered the T-odd effects due to
the CPC unparticle phase
in $B\to K^* \ell^{+} \ell^{-}$, which vanish in the SM as
the  CPV weak phase is negligible in the $b\to s$ transition.
Finally, we remark that the
direct CPA for $B_{d}\to\tau^{+}\tau^{-}$ could be accessible at a
future super-B factory, such as the SuperKEKB which could produce
$O(10^{10})$ $B\bar{B}$ pairs in its initial stage \cite{SuperKEKB}.
Moreover,
4400 events/year for
$B\to K^* \ell^{+} \ell^{-}$ decays
will be produced at the LHCb, corresponding to the
accuracy of the rate asymmetries being around
percent level
\cite{LHCb}. Clearly, we have a great chance to observe the virtual
unparticle phase effects in the $B$ system.\\

\noindent
{\bf Acknowledgments}

We would like to thank Dr. Kai-Feng Chen, Dr. John N. Ng and Dr.
Tzu-Chiang Yuan for useful discussions. This work is supported in
part by the National Science Council of R.O.C. under Grant \#s:
NSC-95-2112-M-006-013-MY2 and NSC-95-2112-M-007-059-MY3.


\end{document}